\documentclass[10pt,conference]{IEEEtran}
\IEEEoverridecommandlockouts

\usepackage{xcolor}
\usepackage{amsmath,amssymb,amsfonts}
\usepackage{algorithmic}
\usepackage[pdftex]{graphicx}
\usepackage{grffile}
\usepackage{textcomp}
\usepackage{xcolor}
\usepackage[numbers]{natbib}
\usepackage{subcaption}
\usepackage{xspace}
\usepackage{booktabs, tabularx}
\usepackage{url}
\usepackage{multirow}
\usepackage[colorlinks=false, linkcolor=blue, citecolor=blue, bookmarks=true,pagebackref=false]{hyperref}
\usepackage{tcolorbox}
\usepackage{verbatim}
\usepackage{balance}
\usepackage{float}
\usepackage[T1]{fontenc}
\usepackage{listings}
\usepackage{makecell}
\usepackage{natbib}[order]

\newcommand{\REM}[1]{}

\definecolor{dkgreen}{rgb}{0,0.6,0}
\definecolor{gray}{rgb}{0.5,0.5,0.5}
\definecolor{mauve}{rgb}{0.58,0,0.82}
\definecolor{dkred}{rgb}{0.6, 0.1, 0}
\definecolor{dkblue}{rgb}{0, 0, 0.6}
\definecolor{lightgreen}{RGB}{205, 255, 216}
\definecolor{lightred}{RGB}{255, 220, 224}
\definecolor{lightblue}{RGB}{179, 217, 255}
\definecolor{brass}{rgb}{0.71, 0.65, 0.26}
\definecolor{violet(ryb)}{rgb}{0.53, 0.0, 0.69}
\definecolor{mediumorchid}{rgb}{0.73, 0.33, 0.83}
\definecolor{darkorange}{rgb}{1.0, 0.55, 0.0}
\definecolor{navyblue}{rgb}{0.36, 0.54, 0.66}
\definecolor{amber}{rgb}{1.0, 0.75, 0.0}

\usepackage[linesnumbered,ruled,vlined]{algorithm2e}
\usepackage{amsmath}

\newcommand{\phead}[1]{\vspace{1mm} \noindent {\bf #1}}

\newcommand{\tool}{{{MobileUPReg}}\xspace}
\newcommand{\rqbox}[1]{\begin{tcolorbox}[left=4pt,right=4pt,top=4pt,bottom=4pt,colback=gray!5,colframe=gray!40!black,before skip=2pt,after skip=2pt]#1\end{tcolorbox}}

\usepackage{hyperref}
\hypersetup{
  pdfauthor={Anonymous},
  pdftitle={},
  pdfsubject={},
  pdfkeywords={}
}

\begin{document}
\title{\tool: Identifying User-Perceived Performance Regressions in Mobile OS Versions}

\author{
    \IEEEauthorblockN{Wei Liu\textsuperscript{1}, 
    Yi Wen Heng\textsuperscript{1},
    Feng Lin\textsuperscript{1},
    Tse-Hsun (Peter) Chen\textsuperscript{1}, 
    Ahmed E. Hassan\textsuperscript{2}}
    \IEEEauthorblockA{\textit{\textsuperscript{1}Software PErformance, Analysis, and Reliability (SPEAR) lab, Concordia University, Montreal, Canada}}
    \IEEEauthorblockA{\textit{\textsuperscript{2}Queen's University, Canada}}

    \IEEEauthorblockA{w\_liu201@encs.concordia.ca,
    he\_yiwen@encs.concordia.ca,
    feng.lin@mail.concordia.ca,
    }
    \IEEEauthorblockA{
    peterc@encs.concordia.ca, 
    ahmed@cs.queensu.ca
    }
}

\maketitle

\begin{abstract}
Mobile operating systems (OS) are frequently updated, but such updates can unintentionally degrade user experience by introducing performance regressions. Existing detection techniques often rely on system-level metrics (e.g., CPU or memory usage) or focus on specific OS components, which may miss regressions actually perceived by users—such as slower responses or UI stutters. To address this gap, we present \tool, a black-box framework for detecting user-perceived performance regressions across OS versions. \tool runs the same apps under different OS versions and compares user-perceived performance metrics—response time, finish time, launch time, and dropped frames—to identify regressions that are truly perceptible to users. In a large-scale study, \tool achieves high accuracy in extracting user-perceived metrics and detects user-perceived regressions with 0.96 precision, 0.91 recall, and 0.93 F1-score—significantly outperforming a statistical baseline using the Wilcoxon rank-sum test and Cliff's Delta. \tool has been deployed in an industrial CI pipeline, where it analyzes thousands of screencasts across hundreds of apps daily and has uncovered regressions missed by traditional tools. These results demonstrate that \tool enables accurate, scalable, and perceptually aligned regression detection for mobile OS validation.
\end{abstract}


\section{Introduction}
\label{sec:introduction}
Mobile operating systems, including Android and iOS, along with vendor-customized Android variants such as MIUI~\cite{MIUI}, One UI~\cite{One_UI}, and ColorOS~\cite{ColorOS}, form the foundation of billions of mobile applications worldwide. Even minor updates to these systems, such as changes to the task scheduler, rendering pipeline, or exposed APIs, can inadvertently degrade app responsiveness or visual smoothness.
We consider such degraded app behavior—particularly when it affects how users perceive system performance—as a form of \textit{performance regression}.
Given the frequency of OS updates, vendors must ensure that each new version does not introduce performance regressions that negatively impact the perceived performance and overall user experience~\cite{2014_software_mobile_complain}.

However, detecting regression across OS versions—particularly those that are actually perceptible to end users—remains a significant challenge in practice. For each OS update (e.g., from version 13.0.12 to 13.0.13), vendors must determine whether the new version degrades app behavior in ways that matter to users, such as longer launch times or slower response. 
Existing tools and metrics often fall short in providing actionable insight into user-facing regressions. 
As a result, vendors typically rely on labor-intensive strategies like manual Quality Assurance(QA), limited A/B rollouts, or post-release feedback—approaches that are expensive, slow, and reactive. This situation highlights the urgent need for an automated, scalable method to detect user-perceived performance regressions before an OS version is shipped.

Detecting user-perceived performance regressions in mobile OS is particularly challenging because of the diverse apps and hardware devices.
Unlike server applications, where applications are centrally hosted and performance metrics (e.g., CPU usage, memory consumption) can be systematically collected, mobile devices are highly distributed and privately owned. 
It is challenging to know the performance impact of an OS update on the hundreds of apps that users install on the device. 
Moreover, collecting system-level metrics at scale requires end-user consent and cooperation, which is often impractical due to privacy concerns and performance overhead on mobile devices. More importantly, these system-level metrics may not align with user-perceived performance issues.

Most existing mobile performance tools, including profiling and instrumentation frameworks, such as ADB, Perfetto, and Android Studio Profiler~\cite{adb, perfetto, Android_Studio_Profiler, 2012_OSDI_AppInsight, 2019_MOBILESoft_PerfProbe, 2022_EMSE_AppSPIN}, and benchmarking suites like AnTuTu~\cite{AnTuTu_benchmark}, Geekbench6~\cite{Geekbench6}, 3DMark~\cite{3DMark}, and PCMark~\cite{pcmark_android}, primarily focus on system- and component-level metrics such as CPU utilization, memory consumption, GPU performance, and storage I/O. While these metrics provide valuable insights into hardware and resource usage, they often fail to capture user-perceived performance during real-world interactions. 
A recent study~\cite{2022_IST_resource_influences_UI_responsiveness} found that high resources often have weak or inconsistent correlations with user-perceived responsiveness. In some cases, applications with high CPU or memory usage can still feel smooth and responsive to users, while others with low resource usage may appear slow and sluggish. To better capture user-perceived performance, recent efforts~\cite{2025_ICSE_SEIP_GUIWatcher, 2025_MobileGUIPerf} turn to analyzing GUI screencasts to capture user-perceived delays, but they primarily focus on individual app behavior rather than system-wide regressions.

To address these limitations, we collaborated with Company A to develop \tool, a black-box, user-centric framework designed to detect performance regressions across OS versions that are perceptible to users. Instead of relying on system-level metrics, \tool runs the same real-world apps under different OS versions, analyzes screencasts of automated GUI interactions, and extracts four user-perceived performance metrics—response time, finish time, launch time, and dropped frames. It then compares these metrics using the Wilcoxon rank-sum test, Cliff's Delta, and a perceptual threshold to detect regressions that are both statistically and perceptually meaningful.
While recent work such as GUIWatcher~\cite{2025_ICSE_SEIP_GUIWatcher} and MobileGUIPerf~\cite{2025_MobileGUIPerf} also analyze screencasts to detect performance issues, they focus on detecting specific GUI issues (e.g., janky frames, frozen frames) or visual delays within a single app. In contrast, \tool is designed to capture system-wide performance regressions by observing how the same app responds under different OS versions.

We design \tool based on three key observations.  
\textbf{First}, users primarily interact with mobile OS through applications~\cite{2022_TSE_Mobile_Performance_Optimization}, and recent statistics show that they spend over 88\% of their smartphone time within apps~\cite{BuildFire}. Therefore, using apps as natural probes to detect performance regressions across OS versions better reflects real-world usage scenarios. 
\textbf{Second}, we detect performance regressions across OS versions by observing degradations in app behavior. As a result, the regressions we uncover are more likely to impact user-perceived performance, such as slower launches, delayed responses, or UI stutters.
\textbf{Third}, apps and their usages are diverse, covering a wide range of real-world scenarios that are often missed by system-level tests based on synthetic workloads. In contrast to these synthetic tests—which typically target specific components, such as I/O stress or CPU-bound workloads—real apps exercise multiple subsystems simultaneously, offering broader and more realistic coverage.
These user-facing issues are particularly important to OS vendors, as they directly affect perceived system quality and user satisfaction.

To evaluate the effectiveness of \tool, we conducted a large-scale study using real-world mobile applications across multiple OS versions in an industrial CI pipeline. \tool accurately extracts user-perceived performance metrics from screencasts, achieving mean absolute errors of 19\,ms for response time, 76\,ms for finish time, 60\,ms for launch time, and 0.2 frames for dropped frames. In detecting user-perceived performance regressions, \tool reaches a precision of 0.96, recall of 0.91, and F1-score of 0.93—outperforming a baseline using the Wilcoxon rank-sum test and Cliff's Delta by 20\% in precision and 8\% in F1-score. \tool has been deployed in production, where it analyzes thousands of screencasts across hundreds of apps on a daily basis. Developers report that it consistently detects regressions missed by system-level tools and provides actionable, perceptually aligned insights that better reflect actual user experience.

The main contributions of this paper are as follows: 
\begin{itemize}

    \item We present \tool, a black-box framework for detecting user-perceived performance regressions across OS versions.
    
    \item \tool extracts four user-perceived performance metrics—response time, finish time, launch time, and dropped frames—with high accuracy, enabling reliable and scalable assessment of regressions across mobile OS versions.
    
    \item \tool achieves 0.96 precision, 0.91 recall, and 0.93 F1-score in detecting user-perceived performance regressions, significantly outperforming a baseline using the Wilcoxon rank-sum test and Cliff's Delta.

    \item \tool has been deployed in an industrial CI pipeline, where it analyzes thousands of screencasts daily and has uncovered regressions missed by system-level tools. It now plays a key role in pre-release OS validation.
\end{itemize}

\phead{Paper organization.} 
Section~\ref{sec:background} introduces the background on user-perceived performance metrics. Section~\ref{sec:related} reviews related work. 
Section~\ref{sec:approach} presents the details of \tool.
Section~\ref{sec:discussion} discusses its industrial deployment and practical impact.
Section~\ref{sec:evaluation} evaluates the accuracy and effectiveness of \tool. 
Section~\ref{sec:threats} outlines threats to validity, and Section~\ref{sec:conclusion} concludes the paper.

\section{Background}
\label{sec:background}

Existing approaches to detecting performance regressions primarily rely on system-level metrics—such as CPU usage, memory consumption, or method execution time. However, high resource usage does not necessarily lead to user-perceived performance degradation, such as slower UI responses~\cite{2022_IST_resource_influences_UI_responsiveness}.
For example, \citet{2022_IST_resource_influences_UI_responsiveness} observe cases where CPU utilization appears normal, yet the UI becomes noticeably unresponsive. They also report that high memory usage and swap usage—even close to 100\%—do not necessarily lead to responsiveness degradation in the UI.
To address this limitation, we detect mobile system performance regressions by running the same mobile apps across different OS versions and comparing their visual behavior. Specifically, we extract user-perceived performance metrics (e.g., response time) from the GUI screencasts—video recordings of an app’s visual feedback to user actions (e.g., tap)—to observe how visual responsiveness changes across OS versions. A regression is flagged when the same app exhibits degraded visual responsiveness—such as increased response time or more dropped frames—under a newer OS version. These detected OS-induced regressions are user-perceived and directly affect the user experience. 

Next, we introduce the concept of screencasts and describe the user-perceived performance metrics extracted from them.

\subsection{Screencasts for Capturing User-Perceived Performance}
\begin{figure*}[htbp]
	\centering
	\includegraphics[width=0.9\linewidth]{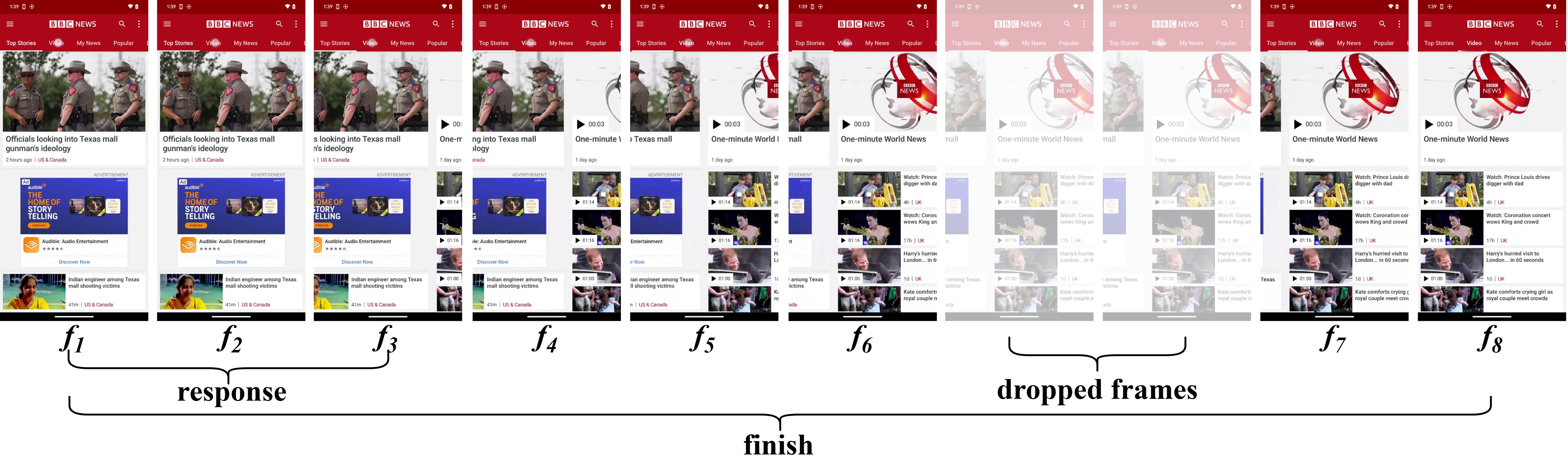}
	\caption{A screencast segment corresponding to a single user interaction (i.e., one screencast per user action). Each frame is annotated by an index (e.g., $f_{1}$). The \textit{response time} spans from $f_{1}$ to $f_{3}$, and the \textit{finish time} from $f_{1}$ to $f_{8}$. Frames between $f_6$ and $f_7$ are semi-transparent to indicate \textit{dropped frames}, where rendering was skipped or delayed.}
	\label{fig_screencast}
     \vspace{-3mm}
\end{figure*}

Since users primarily interact with mobile OS through applications~\cite{2022_TSE_Mobile_Performance_Optimization, BuildFire}, we capture performance from the user's perspective by recording \textit{screencasts}—videos of the device screen—during automated GUI testing. 
Each GUI test script defines a sequence of user actions (e.g., launching an app, tapping a button, switching views), which are executed consistently across OS versions. For each action, we start recording just before execution and stop after a short, fixed wait—allowing the app sufficient time to complete its visual response. This process yields a set of short screencasts, each corresponding to a single \textit{user interaction}. A screencast consists of consecutive visual frames recorded at the device’s native frame rate (e.g., 60 FPS), with each frame annotated by a timestamp. These frame sequences capture how the app visually responds to user input over time, providing a fine-grained, black-box view of system behavior that is independent of source code. As illustrated in Figure~\ref{fig_screencast}, each user interaction is recorded as a short screencast segment, starting just before the action is executed and ending after the app has completed its visual response. The resulting frame sequence reflects how the UI evolves in response to that single user action.

By analyzing such screencasts across OS versions, we can detect subtle regressions in perceived responsiveness that traditional system-level metrics may miss. Screencasts thus enable automated, user-centric performance evaluation at scale in real-world testing environments.

\subsection{User-perceived Performance Metrics from Screencasts}
To detect user-centric regressions, we rely on metrics that reflect how users experience performance during app interactions.
Table~\ref{table:performance_metrics} summarizes commonly used user-perceived performance metrics based on prior human-computer interaction (HCI) and software performance studies~\cite{yan2012fast, 10.1145/1062745.1062747}. \textit{Response time} refers to the time between a user action and the first visible frame update in the GUI. Prior research in HCI suggests users can feel delays in responses that take more than 100\,ms~\cite{10.1145/1062745.1062747}. Hence, even minor delays in GUI updates may thus be perceived as unresponsiveness. 
\textit{Finish time} captures the duration from the user’s action to when the GUI completes all visual transitions (i.e., usually when the last frame stops changing) and becomes ready for the next user interaction. 
\textit{Launch time} is a specific case of finish time, measuring how long it takes to launch an application. Performance issues are more common in \textit{cold starts}~\cite{lin2013addressing}, where the app is launched with no existing process or cached state, as they typically exhibit the worst performance from a user's perspective. \textit{Dropped frames}~\cite{Dropped_frames} are the number of frames dropped during rendering, which may cause visible stutter.

By analyzing frame-by-frame visual changes, we extract four key user-perceived performance metrics (Table~\ref{table:performance_metrics}): \textit{response time}, \textit{finish time}, \textit{launch time}, and \textit{dropped frames}. For example, Figure~\ref{fig_screencast} shows a screencast composed of 8 frames, each labeled with a timestamp indicating its Presentation Time Stamp (PTS)~\cite{FFmpeg_PTS}. For instance, frame 1 ($f_1$) is shown at timestamp 0 ms, followed by frame 2 ($f_2$) at timestamp 16 ms, and so on. In this scenario, the user taps the ``Video'' tab on the screen at $f_1$ to switch to the video page. The mobile system begins responding at frame $f_3$ and completes the transition by $f_{8}$. Several frames between $f_6$ and $f_7$ are dropped, as indicated by the reduced opacity in the figure. As a result, the response time is calculated as $f_3$ - $f_1$, and the finish time is $f_{8}$ - $f_1$.

\begin{table}
\caption{User-perceived performance metric. All metrics are extracted from GUI screencasts and reflect visual feedback as perceived by users.}
\label{table:performance_metrics}
\centering
\scalebox{0.95}{
    \setlength{\tabcolsep}{3pt}
    \begin{tabular}{lp{6.5cm}}
    \toprule
    Metrics & Definition\\
    \toprule
    Response time & Duration from user input to the \textbf{first} visible GUI frame update. \\
    \midrule
    Finish time & Duration from user input to the \textbf{final} GUI frame update. \\
    \midrule
    Launch time & Time taken to launch the application from an initial, non-running state. \\
    \midrule
    Dropped frames & Number of frames dropped during rendering, which may cause visible stutter.\\
    \bottomrule
    \end{tabular}}
     \vspace{-3mm}
\end{table}

\section{Related Work}
\label{sec:related}
In this section, we review prior work related to our study.

\subsection{Performance Regression Detection}
Performance regressions frequently arise during software development \cite{2017_ICSME_Exploratory_Study_of_Performance_Regression}, and many techniques have been proposed to detect them~\cite{2012_ICPE_performance_regressions_statistical, 2020_NSDI_Gandalf, 2022_TSE_PerfJIT, 2024_SOSP_FBDetect, 2024_OSDI_ServiceLab, 2025_ICSE_Early_Detection_of_Performance_Regressions, MobileLab}. These approaches typically compare system-level metrics across software versions—such as test case execution time, server-side response time, throughput, CPU usage, and memory consumption—to identify statistically or practically significant degradations. For example, PerfJIT~\cite{2022_TSE_PerfJIT} uses test execution time to predict performance regressions, while FBDetect~\cite{2024_SOSP_FBDetect} monitors deployment-level changes in system metrics like server response time, CPU usage, and memory consumption.
MobileLab~\cite{MobileLab}, although targeting mobile platforms and tracking metrics like launch time and scroll latency, still relies on instrumentation and system profiling. It does not consider the visual feedback users actually perceive on screen.

In contrast, our approach detects performance regression from the user's perspective on mobile operating systems. We run real mobile apps and assess their responsiveness using user-perceived performance metrics—such as response time, finish time, and frame drops—directly observable from GUI screencasts. This enables us to identify regressions that are visible to users but may be missed by system-level techniques, which is particularly important on mobile platforms, where performance issues are often first perceived through visual feedback rather than system metrics.

\subsection{User-Perceived Performance Issue Detection}
Mobile performance is often assessed through system metrics such as CPU usage, memory consumption, or GPU load. These metrics are widely collected by mobile profiling tools~\cite{adb, perfetto, Android_Studio_Profiler} and dynamic instrumentation tools~\cite{2012_OSDI_AppInsight, 2019_MOBILESoft_PerfProbe, 2022_EMSE_AppSPIN} to monitor resource usage and trace program execution. While effective for internal diagnosis, these metrics often fail to reflect how users actually experience performance. For instance, prior work~\cite{2022_IST_resource_influences_UI_responsiveness} has shown that high resource utilization does not necessarily correspond to visual delays—such as response lag—may occur even when system metrics appear normal. To better capture user experience, we leverage user-perceived metrics—including response time, finish time, launch time, and frame drops—that more directly reflect how users perceive the mobile performance than traditional system metrics which often reflect internal performance.

A few recent studies have also explored user-perceived performance. GUIWatcher~\cite{2025_ICSE_SEIP_GUIWatcher} detects specific GUI lags (e.g., janky frames, long loading frames, and frozen frames) by analyzing mobile screencasts. MobileGUIPerf~\cite{2025_MobileGUIPerf} extracts response time and finish time to identify visual delays. However, both approaches focus on detecting GUI performance issues within individual apps. In contrast, our work differs in both goal and scope. Rather than detecting GUI performance issues within a single app, we systematically measure user-perceived performance across different OS versions using a wide range of real-world apps, enabling broader insights into how system updates affect user experience.

\section{Approach}
\label{sec:approach} 
\begin{figure*}[htbp]
	\centering
	\includegraphics[width=0.9\linewidth]{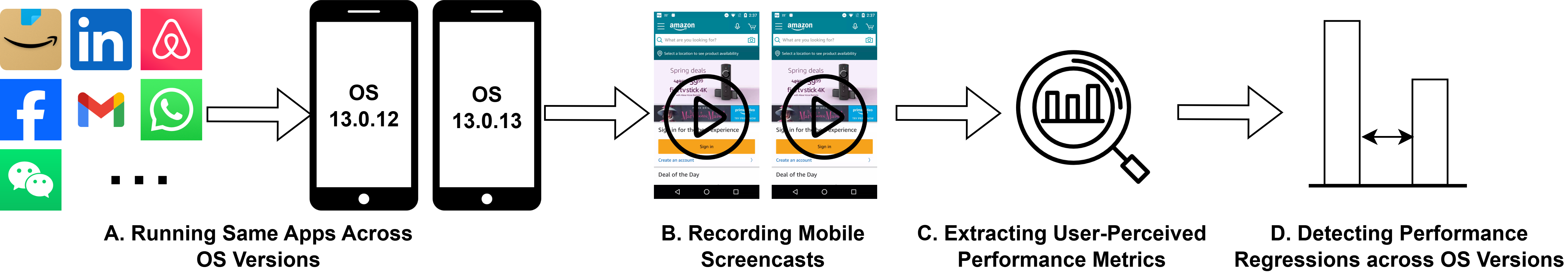}
	\caption{The framework of \tool. The same app versions are executed across different OS versions (e.g., v13.0.12 vs. v13.0.13) under identical conditions.}
	\label{fig_overview_approach}
     \vspace{-3mm}
\end{figure*}

We propose \tool, an automated framework developed in collaboration with Company A, to detect performance regressions introduced by mobile OS updates from the perspective of user-perceived experience. The core idea is to run the same mobile apps (with the same versions) under different OS versions in a controlled environment and compare their observable performance. If a user interaction consistently exhibits degraded performance under a newer OS version, we treat it as a performance regression associated with the OS update—specifically, an app-observable and user-perceived degradation. \tool consists of four main components, as illustrated in Figure~\ref{fig_overview_approach}, : (1) running same apps (and same versions) across mobile OS versions, (2) recording mobile screencasts, (3) extracting user-perceived performance metrics, and (4) detecting performance regressions across OS versions.

This design treats app-level behavior as a practical proxy for underlying OS performance, based on two key insights: mobile apps heavily rely on OS APIs, and app behavior better reflects user-perceived performance than traditional system-level metrics~\cite{ 2022_IST_resource_influences_UI_responsiveness,2025_ICSE_SEIP_GUIWatcher}. Operating entirely in a black-box setting, \tool is particularly well-suited for OS vendors, who typically lack access to app source code but still need to assess the impact of OS changes on real-world app performance.

\subsection{Running Same Apps Across OS Versions}
\label{sec:runningapps} 
To detect user-perceived performance regressions introduced by OS updates, we begin by running a diverse set of top-ranked mobile apps on both pre-update and post-update OS versions. These selected apps span multiple categories and exercise core OS services and resources such as I/O, networking, CPU, and memory, making them suitable proxies for evaluating OS-level performance.

Each app is exercised through carefully designed test scripts that simulate realistic user behavior. These scripts were developed by domain experts at Company A and are designed to (1) cover the app's key features~\cite{2021_ICSE_key_features}, and (2) include interaction patterns that are typically sensitive to performance issues, such as scrolling through content feeds and switching between screens. All scripts are executed automatically and identically across OS versions to ensure repeatability and comparability.  
Note that, since \tool analyzes only the screencasts, our framework does not depend on any specific script design and can operate with any GUI test suite.

To ensure a controlled experimental setting, we hold all variables constant except the mobile OS version. Specifically, we use the same app versions, running them in the same order, and apply the same test scripts on both OS versions at the same time, after the OS update is finalized. This setup helps eliminate time-related variations, such as changes in app backend behavior. To eliminate potential caching effects and ensure consistent initial states, we reinstall each app using the Android Debug Bridge (ADB)~\cite{adb} before executing its test script. Since mobile apps continuously evolve, we periodically update our app set to reflect current usage patterns and maintain representativeness. However, for each comparison across OS versions, we consistently use the same app versions to ensure experimental control. Each interaction is executed 20 times on both the baseline and updated OS versions to account for transient variance and ensure statistical robustness for downstream analysis~\cite{2020_EMSE_statically_detectable_performance_issues, afjehei2019iperfdetector}. This controlled setup helps isolate the performance impact caused solely by the OS update.

\subsection{Recording Mobile Screencasts}
To record app usages with minimal performance overhead, we leverage specialized external hardware: a video capture card. This capture card connects to the mobile device via HDMI, intercepting the display output and performing real-time encoding, thereby offloading the computational workload from the mobile device. We use OBS Studio~\cite{OBS_Studio}, an open-source recording tool running on a separate host machine, to receive the encoded video stream and save it as a video file. Screencasts are recorded at the device’s native frame rate (typically 60 FPS) to ensure that visual changes are fully preserved. 

To enable fine-grained user-centric analysis, we record one screencast per user action. As discussed in Section~\ref{sec:background}, a user interaction is defined as a sequence of frames that starts with a frame in which the user conducts an action (e.g., tapping) and ends just before the next user action. The action occurs a few frames after the first frame. To align the action with the screencast, we match the action's timestamp it occurs with the corresponding timestamp in the video. We wait for a fixed duration (e.g., 10 seconds) after the action to capture its full visual impact. For the implementation of the recording, whenever a user action (e.g., tapping) happens, we start to record and end recording before the next user action. Hence, each screencast consists of sequences of frames that correspond to a single user action, and the first frame is when the user action is conducted. This ensures that each screencast captures only one user input and its resulting visual feedback. As each user interaction is executed 20 times per OS version, we collect 20 screencasts for each interaction under both the baseline and updated OS.

In addition to the screencast videos, we also automatically record metadata during test script execution, including the type of each user actions (e.g., tapping, scrolling, swiping, drawing, or launching via \texttt{adb shell am start}), its screen position (x, y coordinates), the timestamp, and test scenarios. These metadata, together with the screencast, are used in the subsequent offline analysis phase to extract user-perceived performance metrics. The recording pipeline is fully automated and scalable, making it well-suited for large-scale performance testing across OS versions.

\subsection{Extracting User-Perceived Performance Metrics}
After recording the screencasts of user interactions, we analyze them to extract user-perceived performance metrics. As introduced in Section~\ref{sec:background}, we focus on four key metrics: response time, finish time, launch time, and dropped frames. These metrics reflect distinct aspects of the user experience when using the smartphones—responsiveness (response and finish time), app startup latency (launch time), and visual smoothness (dropped frames). Together, they form a comprehensive and practical proxy for evaluating performance from the user's perspective.

\subsubsection{Responsive time and finish time}
Each screencast is a video recording for a single user action (e.g., a tap or scroll), consisting of a sequence of visual frames. Each frame in the video file is internally encoded with a presentation timestamp (PTS)~\cite{FFmpeg}, indicating the exact time it was displayed. To compute \textit{response time} and \textit{finish time}, we first identify three key frames: the \textit{start frame}, when the user action is issued; the \textit{response frame}, when the first visual feedback appears; and the \textit{finish frame}, when the GUI becomes visually stable and ready for next user action. Response time and finish time are computed by subtracting the start frame's timestamp from those of the response and finish frames, respectively.

The start frame is determined based on the timestamp of the user action, recorded by the test script during automated GUI testing. To locate the response and finish frames, we partially adopt the SSIM-based technique from prior work~\cite{2025_MobileGUIPerf}. While the original method detects all three frames using computer vision, we simplify the process by reusing the start frame and focusing only on the visual analysis of the response and finish frames. Specifically, we use the Structural Similarity Index (SSIM) to compare consecutive frames. SSIM values range from 0 (completely different) to 1 (visually identical), capturing structural changes that align with users' visual perception~\cite{2004_SSIM_Transactions_on_Image_Processing}. The response frame is the first frame to show a substantial SSIM drop, while the finish frame is the last frame before SSIM values stabilize.

\subsubsection{Launch Time} Launch time is treated as a special case of finish time, capturing the duration of an app startup interaction. Each screencast corresponds to a user interaction and is labeled as a launch or non-launch based on its associated metadata, which is recorded during the \textit{Recording Mobile Screencasts} step. For launch interactions, we apply the same visual analysis used for finish time—detecting the finish frame via SSIM—to measure the duration from the launch command to the point at which the app becomes visually stable and ready for interaction. This captures the entire startup sequence, including splash screens, loading animations, and startup advertisements.

\subsubsection{Dropped Frames} To detect dropped frames, we analyze the presentation timestamps of consecutive frames within each screencast. If the interval $\Delta t$ between two adjacent frames exceeds the display's vsync threshold (16.67\,ms for 60\,Hz devices), it indicates that one or more frames were dropped during rendering~\cite{UI_jank}. We estimate the number of dropped frames as $\left\lfloor \frac{\Delta t}{16.67\,\text{ms}} \right\rfloor - 1$, and sum these estimates across the interaction to obtain the total number of dropped frames experienced by the user. This metric reflects the smoothness of visual feedback during the interaction.

\subsection{Detecting Performance Regressions across OS Versions}
To detect whether a mobile OS update introduces performance regressions that are noticeable to end users, \tool analyzes user-perceived performance metrics—specifically, response time (RT), finish time (FT), launch time (LT), and dropped frames (DF)—collected from each interaction. Each interaction is executed 20 times under both the baseline and updated OS versions to reduce measurement noise and transient variance. For each metric \( M \in \{\text{RT}, \text{FT}, \text{LT}, \text{DF}\} \), we obtain two sets of measurements: \( T_{\text{base}}^{M} = \{m_1, \dots, m_{20}\} \) and \( T_{\text{updated}}^{M} = \{m_1, \dots, m_{20}\} \).

We compare \( T_{\text{base}}^{M} \) and \( T_{\text{updated}}^{M} \) using the Wilcoxon rank-sum test and Cliff's Delta to assess statistical significance and effect size. However, statistical significance alone does not guarantee perceptibility. 
For example, even if the median finish time increases from 1100\,ms to 1150\,ms across OS versions—yielding a low $p$-value—the change may still remain imperceptible to users and should therefore be given lower priority. 
To ensure that detected regressions are noticeable to users, we introduce a perceptual threshold \( \theta_M \) for each metric: a minimum median difference required to flag a regression. Specifically, a regression is reported only if (1) the updated OS version shows statistically significant degradation and (2) the median difference exceeds a UX-informed threshold, i.e., \( \text{Median}(T_{\text{updated}}^{M}) - \text{Median}(T_{\text{base}}^{M}) > \theta_M \). These thresholds, defined by UX experts from our industry partner, are confidential and not publicly disclosed.  

By combining statistical tests with perceptual thresholds, \tool captures only performance degradations that are both consistent and meaningfully perceptible to end users.

\section{Production Impact and Discussion}
\label{sec:discussion}

\tool's design provides flexibility to accommodate different risk tolerances across development stages. For example, vendors may enforce a strict 0\% regression threshold for stable releases, while allowing higher thresholds for internal testing or experimental builds. By summarizing thousands of fine-grained interaction-level comparisons into a single release decision, \tool enables scalable and customizable regression evaluation tailored to industrial workflows. 
Note that \tool does not attempt to localize regressions to OS internals. It serves as a monitor of user-perceived behavior, offering early signals of performance degradation that affect end users.

While \tool performs regression detection at the interaction level, it supports system-level analysis. By aggregating results across all user interactions. For a given pair of OS versions, \tool computes the proportion of interactions that regress under the new OS version. Developers may define a configurable threshold (e.g., 2\%) to determine whether the overall regression rate is acceptable. If this threshold is exceeded, the OS version is flagged for further review or rollback.

\subsection{A Working Example of \tool}
To demonstrate how \tool functions in practice, we provide an illustrative example of its user interface and workflow. \tool is deployed in production and routinely processes thousands of screencasts generated during automated testing across different OS versions. It supports black-box analysis of app behavior, requiring no instrumentation or access to source code.

\begin{figure*}[htbp]
	\centering
	\includegraphics[width=0.9\linewidth]{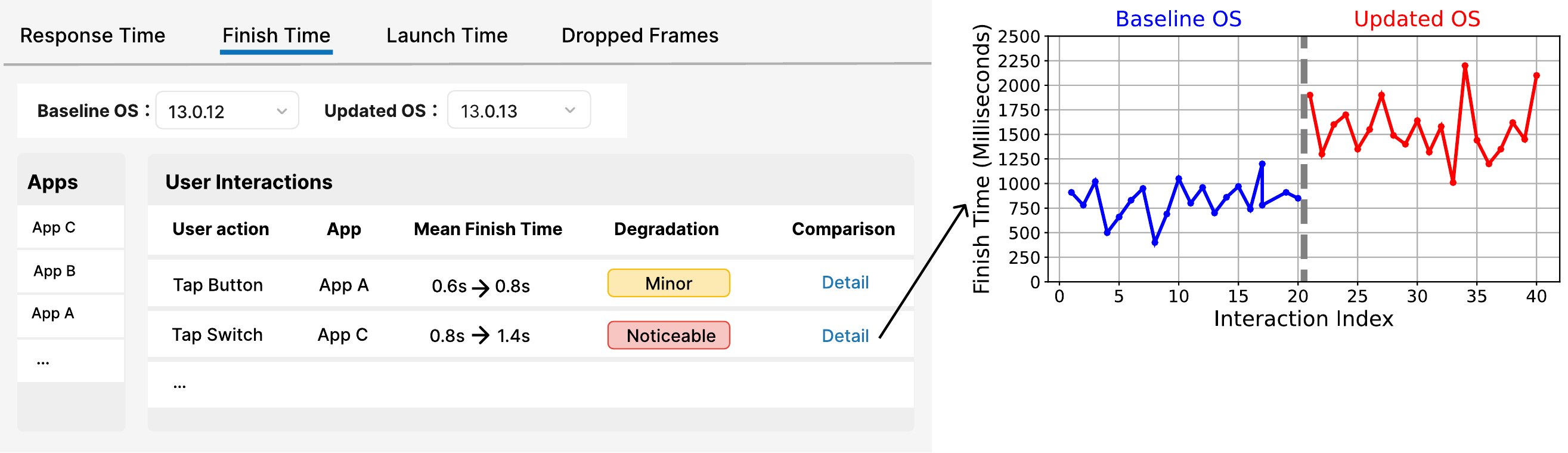}
	\caption{An illustrative interface of \tool.}
	\label{fig_interface}
     \vspace{-3mm}
\end{figure*}

As shown in Figure~\ref{fig_interface}, the main interface allows test engineers to select and compare two OS versions. Once selected, all recorded user interactions are ranked by the severity of performance degradation, based on perception-level transitions (e.g., from ``Excellent'' to ``Noticeable Lag''). Because each interaction is associated with a specific application, the apps themselves are also ranked by their overall impact on degradation, helping developers quickly identify high-risk apps under the new OS version.

By clicking on a particular user interaction entry, developers can drill down into a detailed view. This comparison view displays frame-by-frame differences in visual responsiveness between the two OS versions, along with extracted metrics such as response time, finish time, and dropped frames. Severity indicators highlight whether the metrics exceed perceptual thresholds. This design enables developers to quickly understand how performance regressions manifest visually, assess the severity of each case, and prioritize fixes accordingly.

Due to non-disclosure agreements, we cannot include screenshots of the actual production system. However, Figure~\ref{fig_interface} provides a representative mock-up of the reporting interface to illustrate the typical workflow and insights enabled by \tool.

\subsection{Usefulness of \tool} 
\phead{Explaining User-Perceived Performance Regressions.}
Unlike traditional system metrics (e.g., CPU, memory, or I/O utilization), the regressions identified by \tool are grounded in what end users can actually see and feel—such as delayed responses or stuttered UI transitions. These user-perceived degradations are presented with concrete visual evidence (e.g., side-by-side screencast comparisons), making them highly intuitive and actionable. Developers can immediately understand the nature of the performance issue without interpreting low-level traces or logs. By surfacing perceptual-level regressions tied to specific user interactions and UI elements, \tool helps teams prioritize the issues that are most likely to affect real user experience.

\phead{Insights and Diagnostic Support for Performance Regressions.}
Beyond merely detecting regressions, \tool also provides diagnostic value by aggregating detected issues along two dimensions: the applications involved and the types of user interactions. For example, if a subset of video-related apps consistently shows increased frame drops after a particular OS update, this may suggest regressions in the underlying video decoding subsystem. Similarly, if scrolling-related interactions across multiple apps become slower, developers may suspect problems in the rendering pipeline or gesture handling. This aggregated analysis enables OS vendors and performance engineers to narrow down the potential root causes of regressions and guides further system-level debugging or profiling efforts.

\subsection{Practical Scalability and Flexibility of \tool} 
\tool is designed to be scalable and adaptable across different OS development scenarios, ranging from small-scale updates to major version releases. Small updates typically involve optimizations or new features implemented within a single development branch. In contrast, large-scale OS releases integrate changes from multiple branches, often encompassing extensive modifications across the system. \tool is applicable to both scenarios, enabling either targeted regression checks or comprehensive testing depending on the scope of the update. It's scalability is reflected in three key aspects:

\phead{Parallel Deployment at Scale.}
One challenge we encountered when deploying \tool in production is the time overhead associated with analyzing screencasts, as each regression may involve thousands of screencast videos. To support fast turnaround in industrial deployment, \tool is designed to operate in parallel at both the testing and analysis stages. During testing, we simultaneously execute GUI test cases across multiple mobile devices, each responsible for a portion of the application pool or user interactions. On the analysis side, screencast videos are distributed across multiple host machines for parallel processing. Because metric extraction is fully automated and stateless, the analysis pipeline can easily scale with available computing resources. These engineering strategies make \tool well-suited for large-scale continuous OS testing pipelines.

\phead{Selective GUI Testing for Targeted Analysis.}
In practice, OS updates vary in scope and complexity. For smaller updates—such as those focused on specific subsystems (e.g., app launching)—developers can selectively execute a subset of relevant GUI tests to evaluate only the affected functionalities. For example, if the update involves changes to the launch flow, only GUI tests involving app launches need to be executed. This targeted testing enables the collection of a smaller number of relevant screencasts, significantly reducing testing time and resource consumption while still capturing potential regressions. In contrast, for major OS releases with broader impact, a full suite of GUI tests can be executed to ensure comprehensive coverage. This flexible testing granularity allows \tool to scale with the size and focus of the OS update.

\phead{Configurable Thresholds for Scalable Regression Decisions.}
During regression analysis, \tool provides project managers with customizable control over how performance regressions are interpreted. They can specify tolerable regression thresholds (e.g., allowing up to 2\% of interactions to degrade in performance) or apply stricter criteria for critical applications (e.g., regressions in top-tier apps are unacceptable regardless of the overall rate). This configurable design supports decision-making at scale by balancing regression severity against real-world constraints such as release timelines or app criticality. By allowing flexible threshold settings, \tool adapts to various industrial release workflows, ranging from internal testing builds to public beta and stable releases.

In summary, \tool supports scalable deployment through both adaptive testing strategies and flexible regression evaluation criteria, making it practical for integration into real-world OS development pipelines of varying scope and complexity. It can serve as a key component in pre-release validation workflows, helping OS vendors detect and diagnose performance regressions before public rollout.

\section{evaluation} 
\label{sec:evaluation}
Due to a non-disclosure agreement (NDA) with our industry partner (Company~A), we are unable to release the raw data (e.g., screencasts, apps) used in our evaluation. However, all experiments were conducted on real-world mobile applications selected from internal automated GUI testing pipelines in production environments. We report detailed evaluation results for each research question (RQ1 and RQ2) using representative metrics. While the data itself cannot be shared, our methodology and findings are fully reproducible in similar settings.

\subsection*{RQ1: What is the accuracy of \tool's performance metric extraction?}
\label{sec:rq1}
\phead{Motivation.}
\tool relies on accurately extracting user-perceived performance metrics—such as response time, finish time, launch time, and dropped frames—from screencasts. While prior work~\cite{2025_MobileGUIPerf} has validated parts of the pipeline (e.g., response time and finish time), we extend the metric set to include launch time and dropped frames, offering a more comprehensive view of user experience. Moreover, our industrial deployment utilizes screencasts collected from popular, user-facing apps that better represent real-world OS usage. Hence, we evaluate the accuracy of performance metric extraction in this RQ.

\phead{Approach.}
We randomly sampled 500 launch interactions and 500 non-launch (in-app) interactions from internal automated GUI testing.
To reduce annotation overhead, we randomly selected one execution (out of 20) for each user interaction. This does not affect evaluation accuracy, as all executions follow the same deterministic interaction script.
For each interaction, we first extract all video frames from the screencasts along with their timestamps using \texttt{FFmpeg}~\cite{FFmpeg}. Five user experience experts from Company~A were then invited to independently inspect the frames and annotate the response and finish frames. After the initial annotation, the experts reviewed their results together to resolve discrepancies and reach a consensus. The response time and finish time were then computed based on the timestamps of the annotated frames.

For dropped frames, the ground truth was established from Android system logs collected during GUI testing. The SurfaceFlinger and Choreographer components emit log entries when frames are dropped due to missed rendering deadlines (e.g., ``Skipped N frames'')~\cite{Dropped_frames}. Since both the screencasts and logs record absolute wall-clock timestamps, we aligned each user interaction with its corresponding log segment based on timestamp overlap. We then analyzed these logs and selected only interactions that exhibited at least one dropped frame, as most interactions had none. Including zero-drop cases would trivially reduce the mean absolute error (MAE), so we focused the evaluation on dropped-frame interactions. This filtering yielded hundreds of user interactions for evaluation.

The timestamps extracted by \tool were compared against the ground truth labels to evaluate metric accuracy. For each user interaction, we calculated the absolute error between the extracted value and its ground truth—measured in milliseconds for response time, finish time, and launch time, and in frame counts for dropped frames. The final mean absolute error (MAE) was computed by averaging these errors across all interactions.

\phead{Results.} 
Table~\ref{tab:metric_accuracy} presents the accuracy of user-perceived metric extraction by \tool.  
For response time and finish time, the MAE are 19\,ms and 76\,ms, respectively, which are comparable to or even better than those reported in prior work~\cite{2025_MobileGUIPerf}, despite being evaluated on a more diverse and representative set of mobile apps.
One key reason for this improvement is that \tool does not require explicit segmentation of user interactions from a single screencast, which is known to be error-prone and can significantly affect metric extraction accuracy.  
For example, prior work~\cite{2025_MobileGUIPerf} sometimes merges consecutive interactions due to missed segmentation boundaries, causing the finish frame of one interaction to be mistaken for that of the next—leading to large overestimations.
For launch time (a special case of finish time during app startup), the MAE is 60\,ms, slightly lower than that of general finish time. This is likely because the finish frame of a launch interaction (i.e., app homepage) varies less than that of general user interactions, making it easier to identify. 
In HCI research, perception thresholds such as 100\,ms for response time and 1,000\,ms for finish time are commonly used to assess noticeable performance delays~\cite{1968_AFIPS_Response_time_in_man_computer, 1994_Usability_Engineering, 2015_High_Performance_Android, 2022_TSE_Survey_of_Performance_Optimization}. Compared to these thresholds, the MAE values of time achieved by \tool are relatively small, suggesting that the extracted metrics are precise enough to reflect user-perceived performance with minimal noise.

\begin{table}[t]
\centering
\caption{Accuracy of \tool for extracting user-perceived performance metrics.}
\label{tab:metric_accuracy}
\scalebox{0.95}{
\begin{tabular}{lcc}
\toprule
\textbf{Metric} & \textbf{Ground Truth} & \textbf{Error (MAE)} \\
\midrule
Response time      & Manual annotation     & MAE = 19\,ms \\
Finish time        & Manual annotation     & MAE = 76\,ms \\
Launch time        & Manual annotation     & MAE = 60\,ms \\
Dropped frames     & System logs           & MAE = 0.2\,frames \\
\bottomrule
\end{tabular}}
\vspace{-0.3cm}
\end{table}

For dropped frames, the MAE is 0.2 frames per interaction (computed only on interactions where the ground truth > 0), demonstrating high consistency with Android system logs. Since frame drops typically become perceptible at three or more consecutive losses, this low error level indicates that our detection is accurate enough for practical use.

Overall, \tool accurately extracts user-perceived performance metrics from the screencasts. These results confirm that the extracted metrics are sufficiently reliable for downstream regression detection in industrial settings.

\rqbox{\tool accurately extracts user-perceived performance metrics from screencasts, achieving low mean absolute errors—19\,ms for response time, 76\,ms for finish time, 60\,ms for launch time, and 0.2 dropped frames—enabling reliable quantification of mobile user experience.}
\subsection*{RQ2: Can \tool accurately detect user-perceived performance regressions across OS versions?}
\label{sec:rq2}
\phead{Motivation.}
Not all changes in performance metrics lead to a noticeable impact on user experience. For example, a small increase in finish time (e.g., 50\,ms) across OS versions may not be perceptible to users. To effectively preserve user experience, regression detection must focus on degradations that are \textit{perceptible to users}, particularly those that are visually observable during app interactions. \tool is designed to detect such user-perceived performance regressions, ensuring that reported issues correspond to degradations that users can actually notice.
Once a performance regression is detected, \tool generates a performance bug report containing screencasts from both OS versions, logs, and test details—helping developers visually inspect the issue and prioritize debugging efforts.
In this RQ, we evaluate whether \tool can accurately identify regressions across OS versions that are perceptible to users at the granularity of individual interactions.

\phead{Approach.}
We collected real-world data from the continuously running CI-based testing infrastructure deployed at Company A. From this dataset, we randomly sampled user interaction pairs executed on both baselines. We updated OS versions, covering multiple version combinations. Each interaction was executed 20 times per OS version, resulting in 40 screencasts per interaction pair. Five user experience experts independently reviewed each interaction pair by watching the screencasts to determine whether the user interaction appeared visually degraded when executed on the newer OS version. The annotators focused solely on visual feedback, assessing the overall user experience without reference to specific performance metrics. A majority vote was used to determine the final label, with discussions held to resolve any ambiguous cases. 

For each of the four types of performance regressions, we collected a balanced set of user interaction pairs, with a proportionally equal number of regressed and non-regressed cases. These samples were randomly drawn from a large-scale dataset, and the regressed cases naturally reflected the distribution of degraded metrics in real-world usage. We then compared \tool’s outputs against the ground truth labels to compute precision, recall, and F1-score.

\phead{Baseline method for evaluation.}
We use the Wilcoxon rank-sum test combined with Cliff's Delta as the baseline, a commonly adopted method for detecting performance regressions~\cite{2017_ICSME_Exploratory_Study_of_Performance_Regression,2022_TSE_PerfJIT,2025_ICSE_Early_Detection_of_Performance_Regressions} by comparing two sets of metric values.
While widely used, this approach only captures statistical differences and may not align with user-perceived performance. For example, in a hypothetical case, the baseline OS produced response times of 112, 124, 125, 113, 113, and 111 milliseconds, whereas the updated OS yielded 120, 126, 129, 130, 121, and 120 milliseconds. Although the difference is minor and likely imperceptible to users, the Wilcoxon test reports a significant $p$-value of 0.0043, and Cliff’s Delta equals 1.0. This illustrates that some statistically significant regressions may involve changes that fall below the threshold of human perception and are unlikely to be noticed by users. These cases would therefore have a lower prioritization.  
We did not use percentile-based methods (e.g., P95, P99) in our evaluation. These methods are designed for large-scale online monitoring~\cite{2024_OSDI_ServiceLab,2024_SOSP_FBDetect} with massive request volumes, and are not suitable for GUI testing scenarios with limited executions per interaction.

\phead{Results.} 
Table~\ref{table:regression_accuracy} presents the detection accuracy of \tool compared to the baseline method (Wilcoxon rank-sum test + Cliff's Delta) across four types of user-perceived performance regressions. Overall, the baseline method achieves high recall (average: 0.94) but relatively low precision (average: 0.80). This is because statistical tests are highly sensitive—even small performance degradations that are not perceptible to users can be statistically significant, leading to false positives.

In contrast, \tool significantly improves precision for all regression types while maintaining comparable recall. Notably, it shows substantial gains for regressions involving longer visual delays, such as launch time and finish time. For instance, in detecting launch time regressions, \tool improves precision from 0.69 to 0.94 (a 36\% increase), and for finish time from 0.84 to 0.98 (a 17\% increase). These metrics typical span several hundred milliseconds~\cite{2025_MobileGUIPerf}, making perceptual thresholds in our approach more effective at filtering out subtle variations that users would not notice.  

The improvement for response time is relatively smaller (0.92 to 0.98 in precision), primarily because the baseline method already performs well on this metric. Response times are typically short (e.g., 100–200 ms)~\cite{2025_MobileGUIPerf}, meaning that even minor shifts can be both statistically detectable and perceptible to users. This makes response time regressions inherently easier to identify in a way that aligns with user perception, explaining the baseline's already high precision. Nevertheless, \tool further reduces borderline false positives by focusing on regressions that are both consistent and meaningfully perceived. 

Overall, \tool achieves higher F1-scores than the baseline for all regression types, demonstrating its ability to balance statistical rigor with user-centered relevance. On average, \tool achieves a precision of 0.96 and a recall of 0.91, reflecting a 20\% improvement in precision and a slight decrease of 3\% in recall compared to the baseline (precision 0.80, recall 0.94). These results confirm that \tool is more effective in identifying regressions that are truly perceptible to users, while avoiding over-reporting statistically significant but imperceptible changes.

\begin{table}[t]
\caption{Detection accuracy (Precision, Recall, F1-Score) of \tool and the baseline (Wilcoxon rank-sum test + Cliff's Delta) for four types of user-perceived performance regressions across OS versions. Relative improvements of \tool over the baseline are shown in the average row.}
\label{table:regression_accuracy}
\centering
\scalebox{0.85}{
    \begin{tabular}{lllll}
    \toprule
    Regression Type & Method & Precision & Recall & F1-Score \\
    \midrule
    \multirow{2}{*}{Response time} 
    & Baseline &0.92  &0.92  &0.92  \\
    & \tool     & 0.98 & 0.92 & 0.95 \\
    \midrule
    \multirow{2}{*}{Finish time} 
    & Baseline &0.84  &0.96  &0.89  \\
    & \tool     & 0.98 & 0.90 & 0.93 \\
    \midrule
    \multirow{2}{*}{Launch time} 
    & Baseline & 0.69 & 0.94 & 0.80 \\
    & \tool     & 0.94 & 0.91 & 0.93 \\
    \midrule
    \multirow{2}{*}{Dropped frames} 
    & Baseline &0.73  &0.95  &0.82  \\
    & \tool     & 0.93 & 0.90 & 0.92 \\
    \midrule
    \multirow{2}{*}{Average} 
    & Baseline  & 0.80  &0.94    & 0.86 \\
    & \tool & \textbf{0.96} (+20\%)& \textbf{0.91} ($-$3\%)& \textbf{0.93} (+8\%)\\
    \bottomrule
\end{tabular}}
\vspace{-0.3cm}
\end{table}

\rqbox{\tool identifies user-perceived performance regressions with high accuracy, achieving 0.96 precision and 0.91 recall—outperforming the baseline by 20\% in precision and 8\% in F1-score. It ensures that only perceptually meaningful regressions—those actually felt by users—are detected and reported.}

\section{Threats to Validity}
\label{sec:threats}
One threat is that our approach compares the same apps across different OS versions to identify regressions potentially caused by OS updates. To reduce interference from unrelated factors, we conducted all evaluations in a controlled environment: we kept app versions, test scripts, and execution order fixed, and synchronized test execution across multiple devices within a narrow time window. We also took steps to ensure stable device conditions—such as consistent network connectivity, sufficient battery levels, minimal temperature variation, and limited background activity—but acknowledge that minor fluctuations may still occur during runtime. These measures help isolate the effects of OS version changes as much as possible. However, transient variations in backend response, system load, or hardware state may still influence observed behavior. Since \tool does not aim to localize the root cause within the OS, such variability may introduce noise into the detected regressions. Our evaluation instead focuses on whether the observed regressions are perceptible to users, which is the most relevant aspect for practical performance monitoring.

Another threat concerns the generalizability of our findings. Although we evaluated \tool on a large number of real-world apps and multiple OS versions, the selected dataset may not fully represent the diversity of mobile applications, OS distributions, or device configurations in the wild. In addition, our GUI testing scenarios target typical daily and stress usage patterns, but do not cover long-running or aging scenarios where performance degradation may occur gradually at the app or OS level. Furthermore, due to confidentiality agreements, we only report selected production case studies, which may limit the generalizability of our deployment insights.

A further threat lies in the choice of user-perceived performance metrics. The selected metrics—such as response time, finish time, launch time, and dropped frames—were chosen based on discussions with user experience experts at company A, who confirmed these are currently the most relevant to OS-level performance. However, other metrics (e.g., frame jitter) may also affect user experience but are not yet considered. Future extensions of \tool can incorporate additional metrics as needed, given the modularity of our analysis pipeline.

Finally, \tool has been primarily applied to detecting performance regressions—associated with OS version changes—that manifest in typical mobile applications. However, certain domains—such as mobile games—present additional challenges. While OS updates can still affect their performance, the continuous rendering and complex animations in these apps may require additional user-perceived metrics to capture regressions more accurately. Future work may explore extended metric sets to better support such scenarios.

\section{Conclusion}\label{sec:conclusion} 
In this paper, we presented \tool, a black-box framework for detecting performance regressions across mobile OS versions from the perspective of user-perceived experience. \tool runs the same apps under different OS versions, records screencasts of user interactions, and extracts key user-perceived performance metrics such as response time, finish time, launch time, and dropped frames. To detect regressions, \tool compares the distributions of these metrics using statistical tests, while enforcing perceptual thresholds to ensure that only degradations noticeable to users are flagged.
Our evaluation shows that \tool can accurately extract user-perceived metrics and consistently detect regressions aligned with human perception, achieving high precision and recall. \tool has been deployed in industry and integrated into CI pipelines, supporting large-scale regression detection across thousands of screencasts daily. Overall, \tool provides a scalable and practical solution for uncovering performance regressions across OS versions that directly impact user experience.

\section{Acknowledgement}
\label{sec:ack}
We thank Company A for providing access to their enterprise systems to support this study. The findings and opinions expressed in this paper are solely those of the authors and do not necessarily reflect the views of Company A or its subsidiaries and affiliates. The results presented do not constitute an evaluation of the quality or performance of Company A’s products.

\balance
\footnotesize
\bibliographystyle{IEEEtranN}
\bibliography{IEEEabrv,ref}

\end{document}